\documentclass{article}
\usepackage{cite}
\usepackage{spconf,amsmath,graphicx}
\usepackage{booktabs} 
\usepackage{subfigure}
\usepackage{url}
\usepackage{CJKutf8}
\title{Node-wise Domain Adaptation Based on Transferable Attention for Recognizing Road Rage via EEG}
\name{Xueqi Gao$^{1}$,  Chao Xu$^{1,2}$, Yihang Song$^{1}$, Jing Hu$^{1,2}$, Jian Xiao$^{1}$,Zhaopeng Meng $^{1,3}$}
\address{$^1$College of Intelligence and Computing,Tianjin University,Tianjin, China\\
$^2$Higher Research Institute,University of Electronic Science and Technology of China,Shenzhen,China\\
$^3$Tianjin University of Traditional Chinese Medicine,Tianjin, China\\
\small \{gxq\_000110\}@tju.edu.cn}

\begin{document}
\begin{CJK}{UTF8}{gbsn}
%\ninept
%
\maketitle
\vspace{-4ex}
\begin{abstract}
Road rage is a social problem that deserves attention, but little research has been done so far. In this paper, based on the biological topology of multi-channel EEG signals,we propose a model which combines transferable attention (TA) and regularized graph neural network (RGNN). First, topology-aware information aggregation is performed on EEG signals, and complex relationships between channels are dynamically learned. Then, the transferability of each channel is quantified based on the results of the node-wise domain classifier, which is used as attention score. We recruited 10 subjects and collected their EEG signals in pleasure and rage state in simulated driving conditions. We verify the effectiveness of our method on this dataset and compare it with other methods. The results indicate that our method is simple and efficient, with 85.63\% accuracy in cross-subject experiments. It can be used to identify road rage. Our data and code are available. https://github.com/1CEc0ffee/dataAndCode.git 
\end{abstract}
\vspace{-2ex}
\section{Introduction}
\label{sec:intro}
Rage is very common in the process of driving, and it is easy to cause drivers to change lanes, forcibly overtake or even attack other vehicles. It is considered to be one of the main reasons for traffic accidents. Usually, drivers are able to detect emotional changes, but when a negative emotion is strong, it can be difficult to shake off its effects. Therefore, it is necessary to detect the driver's emotional state, which can provide guidance for subsequent emotional interventions, such as music soothing or safety prompts\cite{bankar2018driving}.

As the research object of emotion recognition, EEG signal has the advantages of high temporal resolution and direct reflection of brain state, therefore received extensive attention\cite{bota2019review}. In addition, with the rapid development of wearable devices, wireless EEG headsets provide signal quality comparable to gel electrodes, and some automobile manufacturer have also developed biometric seat prototypes\cite{izquierdo2018advanced}. With these foundations, we conduct a study on the identification of road rage based on EEG signals.

The biological topology of multi-channel EEG signals is very critical for emotion recognition, and the graph neural network (GNN)\cite{scarselli2008graph} can be used to mine the rich information. For example, \cite{yin2021eeg} applied fused graph convolution (GCNN) to extract graph domain features and combined with long short-term memory neural network (LSTM) to extract temporal correlations.\cite{song2018eeg} proposed a dynamic graph convolutional network (DGCNN), which first initialized the adjacency matrix with a distance function, and then dynamically learned it in the network.However, the spatial location of channels does not represent their functional dependence or degree of correlation\cite{fornito2016fundamentals}. In this study, we compute mutual information (MI) to characterize functional connectivity between channels for initial values of adjacency matrices.

On the other hand, the distribution of EEG data of different subjects varies greatly which reduces the generalization ability of the model. Many methods have been proposed for this, for example \cite{li2018cross} applied the Domain Adversarial Network (DANN).\cite{zhong2020eeg} proposed two regularizers NodeDAT and EmotionDL to better handle this type of problem. However, existing methods mainly align representations extracted from the whole EEG signal across domains, failing to take into account that channels which inter-domain differences are not same. Combining the idea proposed by \cite{wang2019transferable} on the image classification problem, we quantify the transferability of the channels with an entropy function based on the output of the node-wise domain classifier and embed it into the emotion classification task.

Overall, our contributions include:
(1) The EEG signals of drivers in pleasure and rage states are collected, which constitute a valuable data set.
(2) To characterize functional correlation between channels, we initialize the adjacency matrix with MI, and then dynamically learned in the network.
(3) Combining transferable attention with node-wise domain adversarial network, which make the model focuses on domain-invariant representation and improves accuracy.

\begin{figure*}[t]
\centering
\includegraphics[scale=0.26]{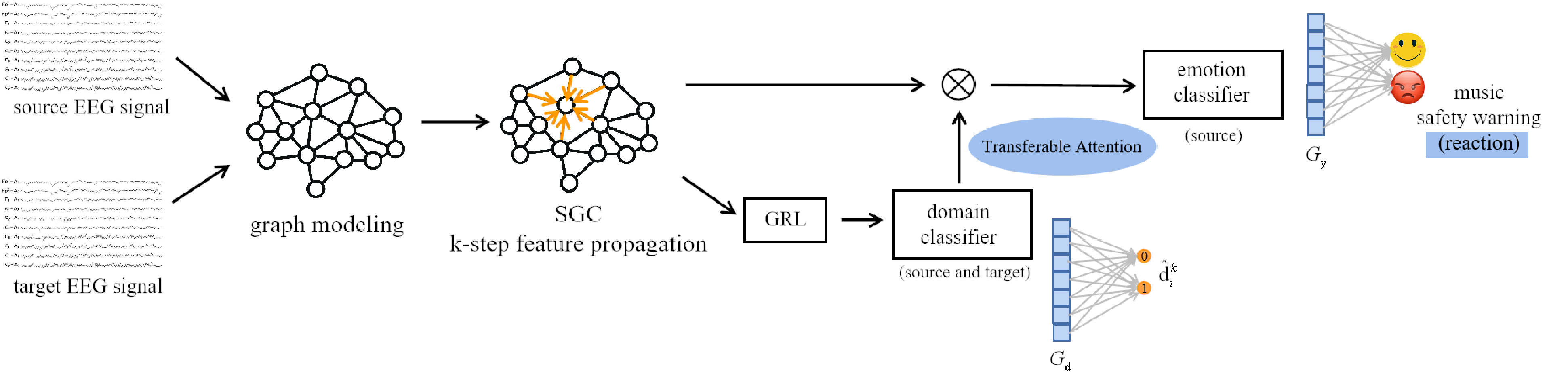}
\caption{The architecture of TA-RGNN in this work}
\label{framework}
\end{figure*}

\section{METHODOLOGY}
\label{sec:format}
In this study, our framework including graph modeling, information aggregation and domain adaptation, as shown in Fig\ref{framework}.

\subsection{Graph Modeling}
Different from images and texts, multi-channels EEG signal belong to non-European domain data, so graph structure modeling is chosen. In this study, the graph is denoted as $G=\left ( V,E \right ) $,$V$ is the set of vertices, represented by the feature matrix $X\in R^{n\times d} $, $n$ denotes the number of channels, $d$ denotes the number of features per channel. $E$ is the set of edges, represented by an adjacency matrix $A\in R^{n\times n} $.

We divide the EEG signal of each channel into five frequency bands($\delta $:0.5\textasciitilde4hz,$\theta$:4\textasciitilde8hz,$\alpha$:8\textasciitilde13hz,$\beta$:13\textasciitilde32hz,$\gamma$:32\textasciitilde50hz). Then, the differential entropy (DE) is extracted to form the feature matrix $X$, and the calculation formula as follows\cite{duan2013differential}:

\begin{equation}
\begin{split}
DE &=-\int_{a}^{b} \frac{1}{\sqrt{2\pi \sigma _{i}^{2} } } e^{-\frac{\left ( x-\mu  \right )^{2}  }{2\sigma _{i}^{2}} }log\left (  \frac{1}{\sqrt{2\pi \sigma _{i}^{2} } } e^{-\frac{\left ( x-\mu  \right )^{2}  }{2\sigma _{i}^{2}} }\right )dx \\
& =\frac{1}{2}log\left ( 2\pi e\sigma _{i}^{2}  \right )  \label{1}
\end{split}
\end{equation}

where $x$ represents an EEG signal of a certain length that approximately obeys a Gaussian distribution $N\left ( \mu ,\sigma _{i}^{2}  \right ) $, $e$ is Euler's constant.

Emotional responses in the brain involve the cooperation of multiple brain regions.We compute the mutual information (MI) between channels as a functional connectivity strength\cite{schlogl2002estimating}, we average over all samples and normalize them for the initial value of the adjacency matrix $A$, as follows：

\begin{equation}
MI\left ( X,Y \right ) =\sum_{x,y}^{}P_{XY} \left ( x,y \right ) log_{2} \frac{P_{XY} \left ( x,y \right )}{P_{X}\left ( x \right ) P_{Y} \left ( y \right )  }  \label{2}
\end{equation}
\begin{equation}
Normalized MI\left ( X,Y \right ) =\frac{2MI\left ( X,Y \right )}{H\left ( X \right )+H\left ( Y \right )  }  \label{3}
\end{equation}

where $P_{XY} \left ( x,y \right ) $ represents the joint probability distribution of the signals $x$,$y$. $P_{X } \left ( x \right )$,$P_{Y } \left ( y\right )$ are the probability distributions of the signals $x$ and $y$. $H\left ( X \right ) $,$H\left ( Y \right ) $ are the entropy of the signals $x$ and $y$.

Negative emotions can activate the right frontal, temporal, and parietal lobes, while positive emotions can activate the left region, forming the spatial characteristics of EEG\cite{davidson1990approach,huang2012asymmetric,davidson1982asymmetrical}. To exploit this information asymmetry, we add several global connections to the adjacency matrix A based on the method proposed in \cite{zhong2020eeg}, setting their initial values as $A_{ij}= A_{ij}-1\in \left [ -1,0 \right ] $, where $\left ( i,j \right ) $ denotes a globally connected pair, including (AF3,AF4), (FC5,FC6), (P7,P8), (O1,O2).

\subsection{Topology-Aware Information Aggregation}
We choose Simple Graph Convolution (SGC)\cite{wu2019simplifying} topology-aware information aggregation of node features. For a given feature matrix $X\in R^{n\times d} $, information aggregation can be expressed as:

\begin{equation}
Z=S^{L} XW\label{4}
\end{equation}

where $S=D^{-\frac{1}{2} } AD^{-\frac{1}{2} }\in R^{n\times n} $, $D$ is the degree matrix of $A$, i.e.$D_{ii} =\sum_{j}^{} A_{ij} $. $W\epsilon R^{d\times d^{'} } $ denotes the weight matrix, and $L$ denotes the number of layers. $Z\in R^{n\times d^{'} } $ is the output, $d^{'}$ denotes the output feature dimension.

\subsection{Focus on Domain-Invariant Representations}
Passing $Z$ into a gradient reversion layer(GRL)\cite{ganin2016domain}, then the domain classifier $G_{d} $ classifies each node representation $reversalZ_{i}^{k} $, the output is $\hat{d} _{i}^{k} = G_{d}^{k } \left (  reversalZ_{i}^{k} \right ) $, representing the probability that the $k$-th node of the $i$-th sample belongs to the source domain. If the domain classifier still cannot distinguish its domain after a certain training, that is, the value of $\hat{d} _{i}^{k} $ is around 0.5, which means that it is a domain-invariant representation, and the $k$-th node can be transferred across domains, which should produce a larger attention value.

The entropy function is defined as $H\left ( p \right ) =-\sum_{j}^{} p_{j} \cdot log\left (   p_{j}\right ) $, which can be used to quantify the transferability of nodes in our study. Specifically, the more transferable the node, the larger $H\left ( \hat{d} _{i}^{k}  \right ) $, and vice versa. According to \cite{wang2017residual}, the effect of false attention can be mitigated by adding residual connections. Before emotion recognition, the node representation is transformed into: 

\begin{equation}
f_{i}^{k} =\left ( 1+H\left ( \hat{d} _{i}^{k}  \right )  \right ) Z_{i}^{k} \label{5}
\end{equation}

In this way, larger weight is given to nodes with small differences between domains, so that these nodes are paid more attention in emotion classification, and the accuracy of the model in cross-subject experiments is improved.

\subsection{Loss Function}
Our model aims to minimize the following loss function:

\begin{equation}
\begin{split}
    Loss &= L_{cls} -\lambda L_{domain} \\
    &=\frac{1}{n_{s} }\sum_{x_{i} \in D_{s} }^{} L_{cls}
\left ( G_{y } \left ( f_{i}  \right ),y_{i}  \right ) +  \alpha \left \| A \right \|  _{1} \\
& -\frac{\lambda }{n_{s} +  n_{t} } \frac{1}{n} \sum_{k=1}^{n} \sum_{x_{i\in D_{s}\cup D_{t}  } }^{} 
L_{domain}\left ( G_{d}^{k}  \left ( reversalZ_{i}^{k}  \right ) d_{i}^{k} \right )  \label{6}
\end{split}
\end{equation}

where $D_{s} =\left \{ \left ( x_{i}^{s} ,  y_{i}^{s}\right )  \right \} _{i=1 }^{n_{s} } $,$D_{t} =\left \{ \left ( x_{i}^{t}\right )  \right \} _{i=1 }^{n_{t} } $ denote the source domain and the target domain, respectively. $x_{i}$ is a sample, $y_{i}$ is the corresponding label. $G_{y} $ denotes emotion classifier and $G_{d} $ denotes domain classifier. $\left \| \cdot  \right \| _{1} $ denotes the $l_{1}$-norm, $\alpha $ and $\lambda $ are hyperparameters.

\section{EXPERIMENTS}
To investigate effective representations of rage states, we carefully design a driving experiment to collect EEG signals.

\textbf{Subjects:} In this study, 10 right-handed college students (6 males, 4 female, mean: 23.90, standard deviation: 1.72) aged 21 to 27 with practical driving experience volunteered for the experiment. Subjects did not have any mental illness, nor did they take drugs, alcohol or caffeine, and ensured reasonable rest and a stable state of mind before the experiment.

\textbf{Experiment Protocol:} For safety reasons,we choose to use the City Car Driving software to simulate in the laboratory environment, and set the steering wheel, joystick, pedals and other equipment to restore the driving experience as much as possible. Previous studies have shown that virtual reality-based scenarios are effective in inducing driving emotions\cite{yan2018induction}, that is, our simulation experiments are simple, safe, and able to achieve the desired effect.

\begin{figure}[h]
\centering
\includegraphics[scale=0.25]{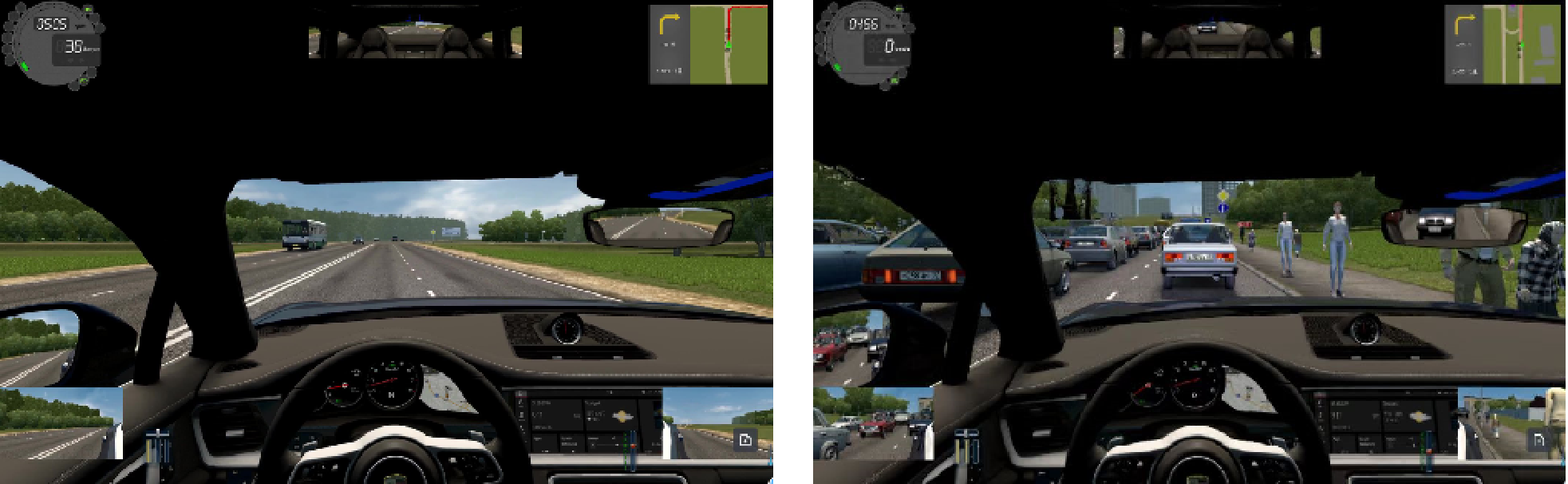}
\caption{Pleasure and rage scenes}
\label{cj}
\end{figure}

We set pleasure and rage scenes separately in the virtual driving software. Among them, the pleasant scene simulates a wide and smooth suburban road with only a small number of cars, and no emergency situation occurs. The rage scene simulates the streets of a commercial area, where traffic is congested, frequently waits for traffic lights, and is also affected by the incorrect behavior of surrounding vehicles and pedestrians interference, as shown in Fig\ref{cj}. In addition, all the scenes are set on a sunny day in summer, in a Porsche Panamera Turbo with automatic transmission, and the driving route is free.

\begin{figure}[h]
\centering
\includegraphics[scale=0.14]{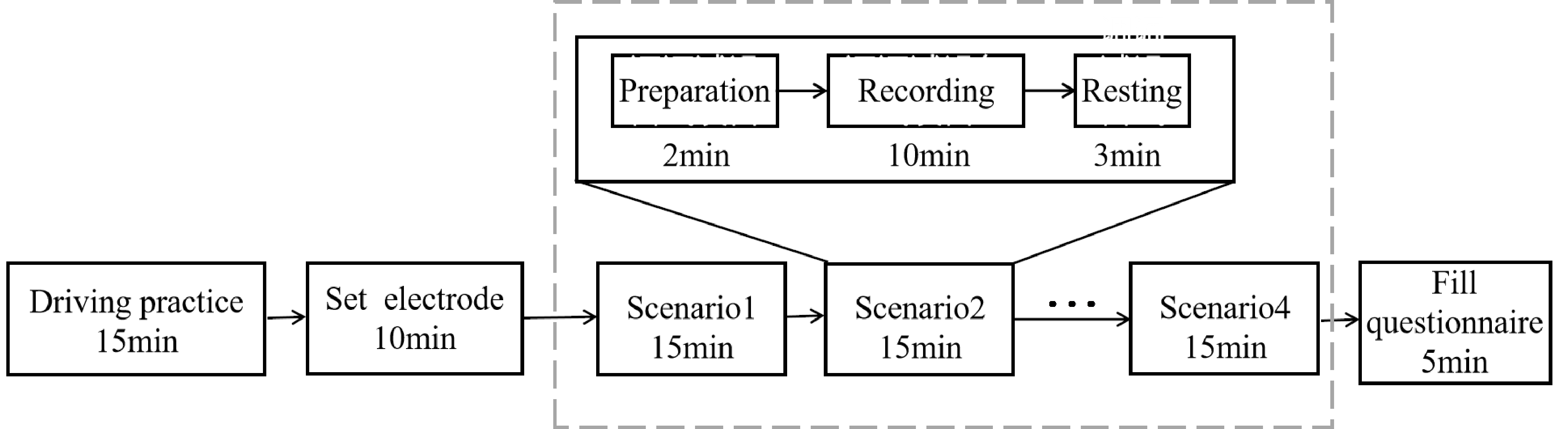}
\caption{Timeline of the experiment}
\label{time}
\end{figure}

The experiment consists of four steps, as shown in Fig\ref{time}. Since the subjects had no experience with driving simulation equipment, they were asked to practice driving before the experiment. Then, the EEG signals were collected while driving under the specified scene, and the experiment was set up to alternate between pleasure and rage scenes, a total of 4 groups. After driving, the subjects filled out a subjective assessment questionnaire, including the emotional state and its intensity and pleasure in each scene. Each experiment lasted about 90 minutes and was performed in an isolated and quiet room.

\begin{figure}[h]
\centering
\includegraphics[scale=0.2]{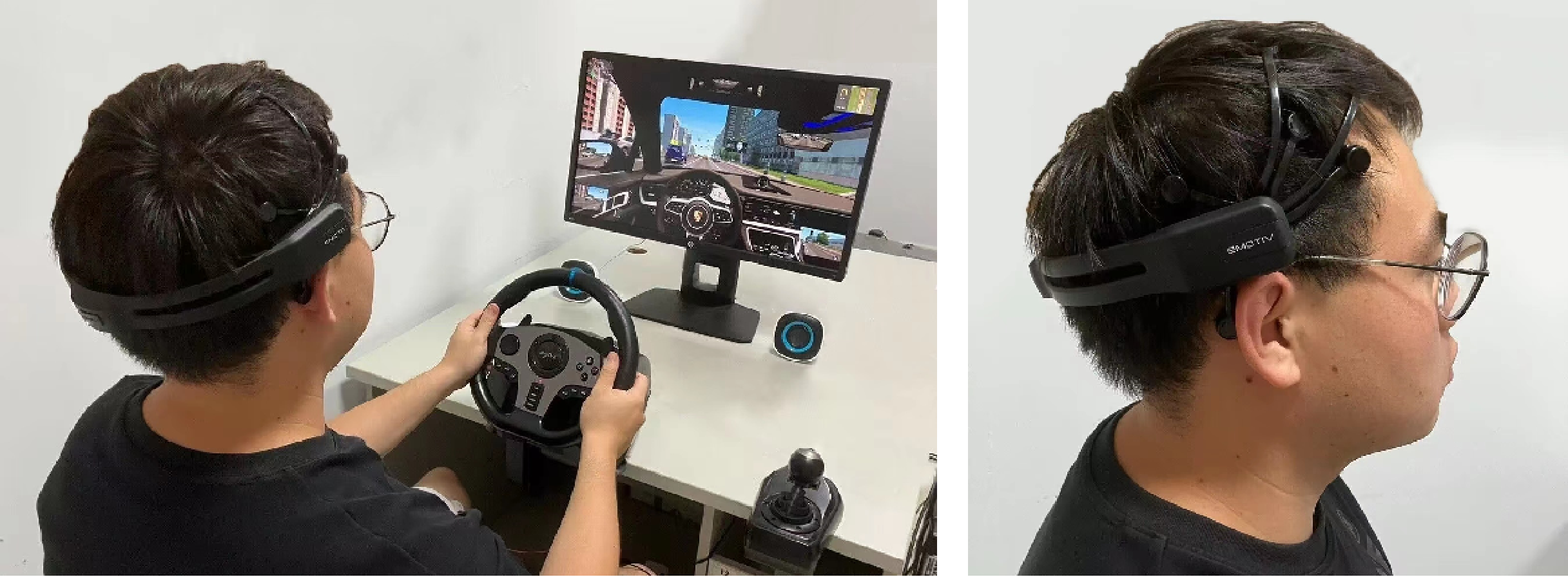}
\caption{Experimental setup}
\label{qzx}
\end{figure}

\textbf{Data Acquisition and Preprocessing:} In order to better fit the proposed application scenario, the Emotiv EPOC+ portable EEG acquisition device is selected for the experiment, and its performance is comparable to that of professional equipment\cite{tyagi2012review}. According to the standard international 10/20 system, 14 electrodes are distributed at key points on the scalp with positions marked AF3, F7, F3, FC5, T7, P7, O1, O2, P8, T8, FC6, F4, F8 and AF4. Additionally, two reference electrodes are located at P3 and P4. Hydrate the sensors with saline prior to acquisition to increase conductivity. During data collection, subjects were advised to limit unnecessary physical movement as much as possible. The experimental setup is shown in Fig\ref{qzx}.

After data collection, the sampling rate was reduced to 128 Hz, and the EEG was band-pass filtered at 0.5\textasciitilde50 Hz to remove voltage drift and high-frequency noise. The EEG was then visually inspected, and heavily contaminated fragments were manually removed. Divide the EEG signals into the same length 1s-epoch with 0.5s overlap. Since Emotiv EPOC+ does not have dedicated EOG, EMG and ECG channels, we implement the FORCe method which is based on a combination of wavelet decomposition, independent component analysis(ICA) and thresholding, capable of removing the aforementioned artifacts. The main steps as follow:

(1) Perform wavelet decomposition on each channel of the EEG signals.

(2) Perform ICA on the approximate coefficients, and exclude ICs containing artifacts according to the differences in AMI, peak value, kurtosis, power spectral density and other characteristics between the artifact and the clean EEG signal.

(3) Calculate the peak area coefficient and perform soft threshold processing.

(4) Reconstruction to obtain a clean EEG signal.

It is experimentally verified that FORCe outperforms the state-of-the-art LAMIC and FasterIC in both offline and online tasks, refer to \cite{daly2014force} for a more detailed description.

\section{RESULTS}
In this section, we introduce the results of TA-RGNN model on the experimental dataset.

\textbf{Performance comparison:} We use the leave-one-subject-out cross validation method to evaluate model performance and compare with previous methods. Table\ref{tabel1} summarizes our experimental results. In \cite{jin2017eeg}, the inter-domain discrepancy is reduced by minimizing the maximum mean discrepancy (MMD) between deep features. \cite{li2018cross} align representations extracted from the entire EEG signal across domains. \cite{zhong2020eeg} implements a more fine-grained cross-domain transfer from the node level. On this basis, we further quantifies node transferability and embeds it into the following emotion recognition task. From Table\ref{tabel1}, we can see that TA-RGNN model achieves a higher accuracy and tends to converge in fewer epochs, which has practical application prospects.

\begin{table}[t]
\begin{center} 
\caption{Comparison with existing work: accuracy, standard deviation and epochs required for model convergence}
\begin{tabular}{cccc}
\toprule
\textbf{model} & \textbf{accuracy} & \textbf{std} & \textbf{epochs} \\
 \midrule
DAN\cite{jin2017eeg} & 69.34 & 11.73 & 90 \\
DANN\cite{li2018cross} & 76.68 & 09.54 & 120 \\
RGNN\cite{zhong2020eeg} & 82.12 & 08.21 & 100 \\
\textbf{TA-RGNN(our work)} & \textbf{85.63} & \textbf{08.79} & \textbf{75} \\
\bottomrule
\label{tabel1}
\end{tabular}
\end{center}
\end{table}

Besides, to verify the effectiveness of each module, according to \cite{zhong2020eeg,shuman2013emerging}, we initialize $A$ with the inter-electrode distance function, and conduct experiments. The emotion recognition accuracy is 83.37\% and 84.05\%, respectively. This shows that multiple brain regions are related to each other when the brain produces emotions, but the spatial location of electrodes does not reflect this connection well.

\begin{figure}[h]
\centering
\includegraphics[scale=0.27]{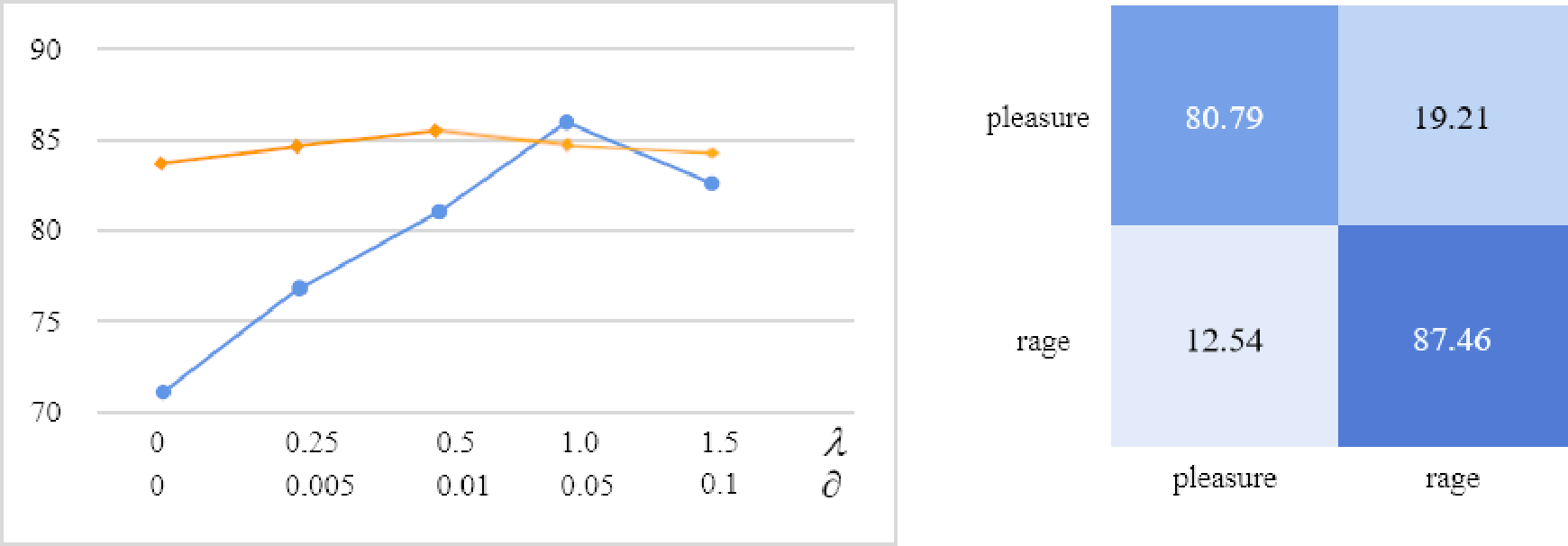}
\caption{Classification accuracy(\%) of TA-RGNN with varying hyper-parameters and confusion matrix}
\label{res}
\end{figure}

\textbf{Sensitivity Analysis:} We set multiple values for hyperparameters for comparison, as shown in Fig\ref{res}. When $\lambda \in \left [ 0,1 \right ]  $, the accuracy rate gradually increases, indicating that the confrontation between the emotion classifier and the domain discriminator enables the network to learn emotion-related, domain-irrelated features. As $\lambda $ continues to increase, it will degrade performance because the learned features lose emotional discrimination.

However, model performance does not change much as hyperparameter $\alpha$ changes. We set $\alpha =0.01$ to avoid strong regularization, according to \cite{zhong2020eeg}.

We plot the confusion matrix as shown in Fig\ref{res}. Compared with pleasure, the model correctly identified a higher proportion of rage, indicating that the EEG signals generated by the subjects in the rage state were more similar.

\vspace{-4ex}
\section{CONCLUSION}
In this study, we propose a model that combines transferable attention and node-wise domain adversarial networks, and demonstrate its effectiveness on cross-subject emotion recognition task. To the best of our knowledge, this is the first study to identify road rage using deep learning. Our approach shows that computing MI and dynamically learning it in the network can better represent the intrinsic relationships between EEG channels. A domain-adaptive model based on transferable attention can focus on domain-invariant representations and improve the accuracy of cross-subject experiments with fewer epochs. In future work, we will explore how to reduce the interference of head shaking on the EEG signal, which will help the model generalize to practical.

\label{sec:refs}
\bibliographystyle{IEEEtran}
\bibliography{reference.bib}
\end{CJK}
\end{document}